\documentclass{appolb}
\usepackage{epsfig}
\usepackage{amssymb,amsmath,bm,bbm}

\begin{document}
\title{Predictions for Flavour Observables in a RS Model with Custodial Symmetry
\thanks{Presented at the Flavianet topical workshop \it{Low energy constraints on extensions of the Standard Model}}
}
\author{Bjoern Duling
\address{Physik Department, Technische Universit\"at M\"unchen, 85748 Garching, Germany}
}
\maketitle
\begin{abstract}
In these proceedings we present the main results for particle-antiparticle mixing and rare decays in the Randall-Sundrum (RS) model with custodial symmetry. To investigate
the strong bound on the Kaluza-Klein (KK) mass scale $M_\text{KK}\gtrsim\mathcal O(20)\,\textrm{TeV}$ implied by the measurement of $\epsilon_K$ we perform a fine-tuning analysis
that on the one hand confirms the quoted bound on the KK mass scale
but on the other hand reveals that consistence with experiment can still be achieved
for small or moderate fine-tuning. In our analysis of rare decays of $K$ and $B$ mesons we find that due to the custodial symmetry the coupling of the $Z$ boson to right handed quarks yields the dominant contribution. This feature of the model leads to distinct patterns and correlations 
that 
allow to distinguish the model from other frameworks of physics beyond the Standard Model (SM).
\end{abstract}
\PACS{11.10.Kk, 12.15.Ji, 12.60.-i, 13.20.Eb, 13.20.He }

\section{Introduction\label{sec:Introduction}}
The extradimensional models first proposed by Randall and Sundrum \cite{Randall:1999ee} are based on a slice of $AdS_5$ that is bounded by the so-called UV and IR (3-)branes and is characterised by a 5-dimensional bulk metric $ds^2=e^{-2ky}\eta_{\mu\nu}dx^\mu dx^\nu-dy^2$.
Effective energy scales in this setup depend on the position along the extra dimension via the warp factor $e^{-ky}$,  which addresses the gauge hierarchy problem if the Higgs field is localised close to or on the IR brane and the geometric model parameters are chosen properly. The only free parameter coming from space-time geometry then is given by 
the mass scale of the lightest KK excitations present in the model. 
Allowing the SM fields except for the Higgs boson to propagate into the 
bulk \cite{Gherghetta:2000qt,Chang:1999nh,Grossman:1999ra} in addition allows to address the flavour problem by naturally generating the hierarchies in the fermion spectrum and mixing angles \cite{Gherghetta:2000qt,Grossman:1999ra}. This is achieved by a non-universal localisation of fermions with different flavours along the extra dimension. In this setup the fermion masses and mixing angles arise from the overlap of the fermion profiles with the IR brane.
Since the localisation of the fermion profiles depends exponentially on flavour dependent $\mathcal O(1)$ bulk mass parameters 
large hierarchies in the fermion spectrum can be achieved in a natural way. On the other hand the non-universal localisation of fermions leads to flavour changing neutral current (FCNC) interactions of the SM fermions and KK gauge bosons, albeit suppressed by the so-called RS-GIM mechanism \cite{Huber:2003tu,Agashe:2004cp} that is an intrinsic feature of the model.

The particular model we will analyse 
is based on an enlarged bulk gauge group 
$G_\text{bulk}=SU(3)_c\times SU(2)_L\times SU(2)_R\times U(1)_X\times P_{LR}$. The fermions in this model are embedded into representations of $G_\text{bulk}$ in such a way that the $T$ parameter \cite{Agashe:2003zs,Csaki:2003zu} and the $Zb_L\bar b_L$ coupling \cite{Agashe:2006at} are protected from too large corrections.
This 
allows for KK masses of order $M_\text{KK}\simeq(2-3)\,\textrm{TeV}$ which are in the reach of the LHC \cite{Cacciapaglia:2006gp,Contino:2006qr,Carena:2007ua,Djouadi:2006rk,Bouchart:2008vp}. A more detailed description of the model can be found e.g.~in \cite{Albrecht:2009xr}.

In the following section we present our main results for $\Delta F=2$ and $\Delta F=1$ processes. A detailed analysis of these processes can be found in \cite{Blanke:2008zb, Blanke:2008yr}. In Sec.~\ref{sec:Conclusions} we present our conclusions.
\section{Analysis of particle-antiparticle mixing and rare decays\label{sec:Analysis}}
\subsection{Particle-antiparticle mixing}
The mixing of neutral mesons imposes very stringent constraints on the KK mass scale of the RS setup \cite{Csaki:2008zd}. In particular $\epsilon_K$, the measure of indirect CP violation in $K^0-\bar K^0$ mixing, sets a lower bound $M_\text{KK}\gtrsim\mathcal O(20)\,\textrm{TeV}$ on the KK mass scale if textureless, anarchic Yukawa matrices are assumed. 
This 
has two reasons:
In $\epsilon_K$ the accidental suppression of the SM prediction by the very small phase of $\left(M^K_{12}\right)_\text{SM}$ meets the presence of effective operators that are induced by the tree level exchange of heavy KK gluons and that are chirally and QCD enhanced by a total factor of approximately 140. Both effects taken together are too large to be overcome by the RS-GIM mechanism; however, all other observables for which either the strong enhancement of the RS contribution or the suppression of the SM prediction is absent are adequately protected.

\begin{figure}
\begin{minipage}{\textwidth}
\begin{center}
\includegraphics[width=.41\textwidth]{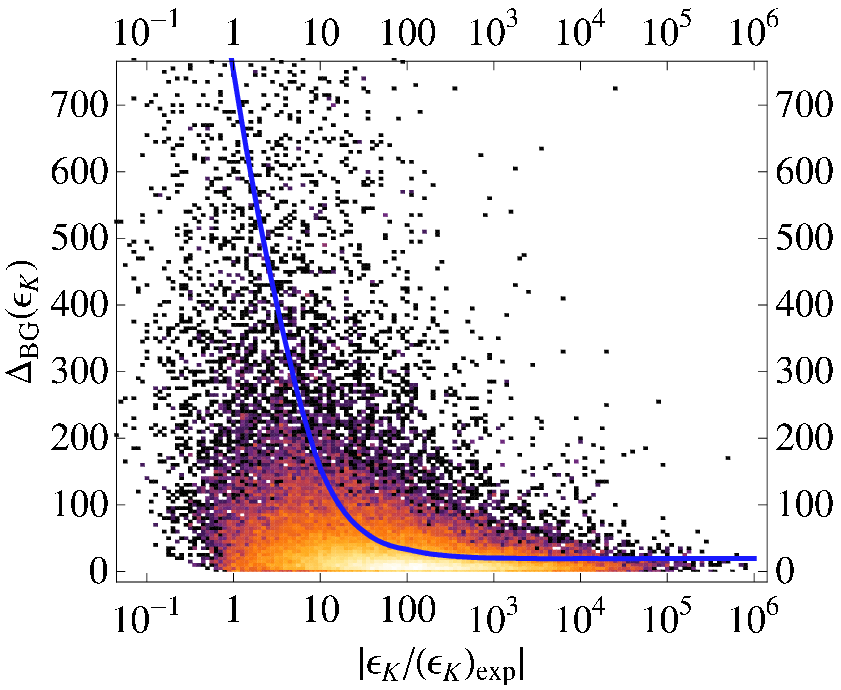}
\hspace{.75cm}
\includegraphics[width=.4\textwidth]{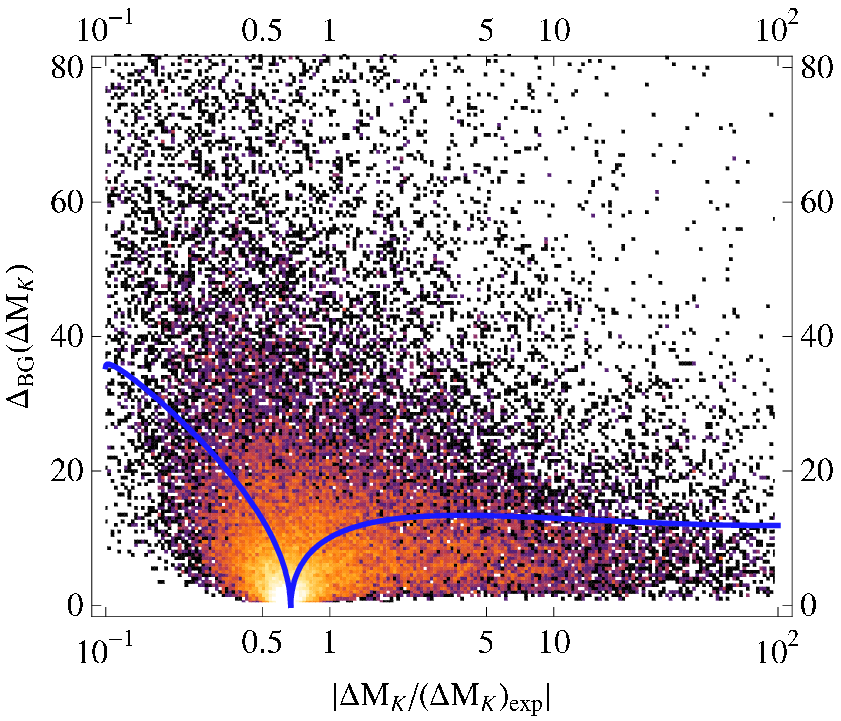}
\end{center}
\end{minipage}
\caption{Left: The fine-tuning $\Delta_\text{BG}(\epsilon_K)$ plotted against $\epsilon_K$ normalised to its experimental value. For visualisation the blue line gives the average fine-tuning. Right: The same for $\Delta M_K$.\label{fig:fine-tuning}}
\end{figure}

To further investigate this matter, in \cite{Blanke:2008zb} (see also \cite{Duling:2009sf}) we perform a fine-tuning analysis for $\epsilon_K$. Our result is given in the left panel of Fig.~\ref{fig:fine-tuning} where we show the fine-tuning in $\epsilon_K$ as a density plot for a KK mass scale $M_\text{KK}\approx2.45\,\textrm{TeV}$. 
It turns out that the RS contribution to $\epsilon_K$ for the chosen KK mass scale is on average by two orders of magnitude too large.
Still, as can also be seen from Fig.~\ref{fig:fine-tuning}, there is a non-vanishing number of points in parameter space that reproduce $\epsilon_K$ with merely small or moderate fine-tuning. This discovery does not remove the tension pointed out in \cite{Csaki:2008zd} - our detailed numerical calculation in fact roughly confirms the bound on the KK mass scale derived therein. However, since due to the reasons outlined above $\epsilon_K$ is the only critical $\Delta F=2$ observable in the RS model (for comparison, in the right panel of Fig.~\ref{fig:fine-tuning} we also show the fine-tuning in $\Delta M_K$ which is small or moderate for all points in parameter space) we proceed with the subset of phenomenologically valid parameter points and investigate their impact on rare decays of $K$ and $B$ mesons.
\subsection{Rare $K$ and $B$ decays}
Using only parameter points that reproduce the experimentally measured $\Delta F=2$ observables with small or moderate fine-tuning we will now have a look at the impact on rare decays \cite{Blanke:2008yr} (see also \cite{Gori:2009tr}). Also these processes receive tree level contributions this time dominantly from the exchange of $Z$, $Z_H$, $Z^\prime$ gauge bosons with masses $m_Z$, $M_\text{KK}$, $M_\text{KK}$, respectively. $Z$, $Z_H$, $Z^\prime$ are the mass eigenstates
of the system containing all $SU(2)_L\times SU(2)_R\times U(1)_X$ gauge bosons with masses up to $M_\text{KK}$\footnote{Higher KK excitations contribute only subleading effects and are neglected in the present analysis.}. All three of those mass eigenstates comprise admixtures of KK gauge bosons and hence have flavour violating couplings to SM quarks. Of these the $Zd_L^i\bar d_L^j$ and $Z^\prime d_L^i\bar d_L^j$ couplings are suppressed by virtue of the custodial symmetry by factors of $\mathcal O(10^2)$ and $\mathcal O(10)$, respectively. In an effective theory analysis \cite{Buras:2009ka} (see also \cite{delAguila:2000kb}) it is shown analytically that the mixing of SM quarks with heavy KK quarks after EWSB yields no significant contribution to the suppressed couplings.

Considering the impact of the custodial symmetry on the quarks couplings of $Z$ and $Z^\prime$, the different masses of the three gauge bosons and the volume suppression in the leptonic coupling of $Z^\prime$ and $Z_H$ one finds that the coupling of the $Z$ boson to right-handed down quarks dominates the RS contribution to rare $K$ and $B$ decays. This surprising result suggests a pattern in rare decays that is very different from the one expected in the minimal RS model without custodial protection. In particular, the rather soft hierarchy in the right-handed $Z$ couplings to different quark flavours
can only partly compensate the hierarchy in the SM CKM factors $\lambda_t^K\ll\lambda_t^d\ll\lambda_t^s$, resulting in much larger relative effects in rare $K$ decays than in decays of $B_d$ or $B_s$ mesons. This is in contrast to the minimal RS model
where the relative NP effects are expected to be similar in size in the $K$, $B_d$ and $B_s$ systems. In the following, the consequences of this finding will be highlighted by means of showing numerical results for the rare decays $K\to\pi\nu\bar\nu$, $B_s\to\mu^+\mu^-$ and $K_L\to\mu^+\mu^-$.\\

\begin{figure}
\begin{minipage}{\textwidth}
\begin{center}
\includegraphics[width=.41\textwidth]{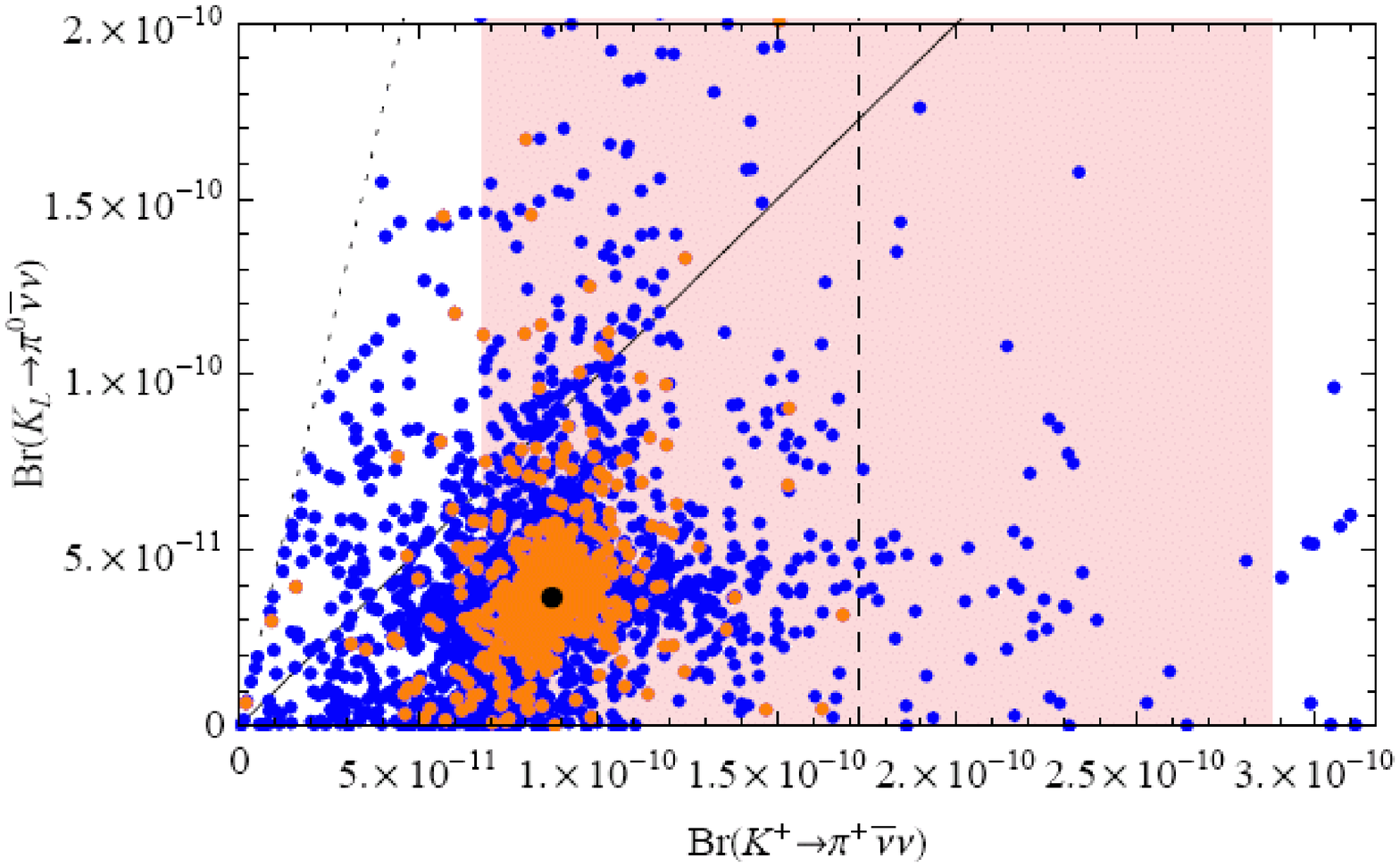}
\hspace{.75cm}
\includegraphics[width=.4\textwidth]{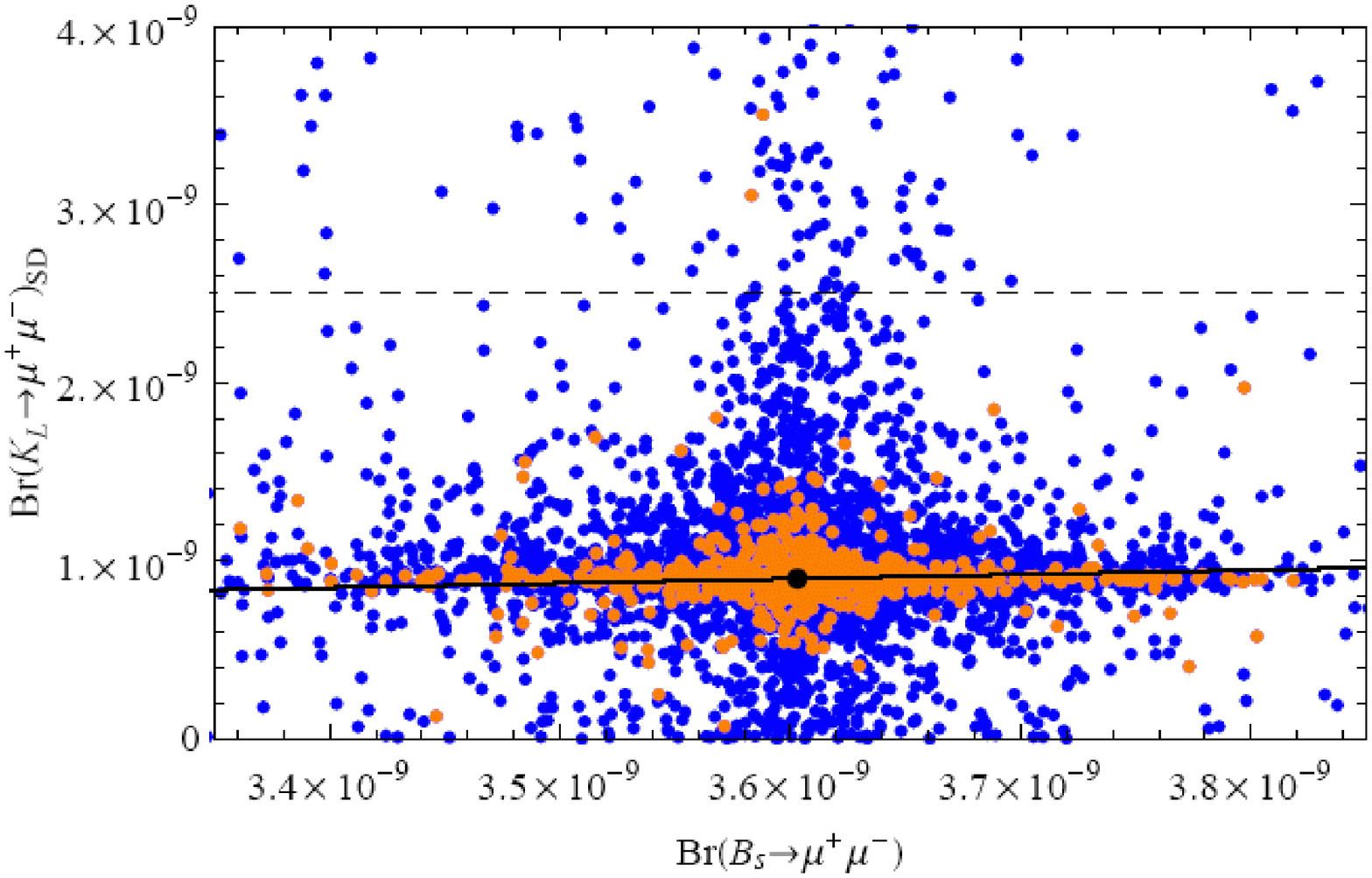}
\end{center}
\end{minipage}
\caption{Left: $Br(K_L\to\pi^0\nu\bar\nu)$ plotted against $Br(K^+\to\pi^+\nu\bar\nu)$. The shaded area represents the experimental $1\,\sigma$-range, the Grossman-Nir bound is displayed by the dashed line. The black point represents the SM prediction. Right: The short-distance contribution to $Br(K_L\to\mu^+\mu^-)$ plotted against $Br(B_s\to\mu^+\mu^-)$.\label{fig:Kpinunu}}
\end{figure}

In the theoretically very clean but experimentally challenging $K\to\pi\nu\bar\nu$ system we find the branching ratios $Br(K_L\to\pi^0\nu\bar\nu)$ and $Br(K^+\to\pi^+\nu\bar\nu)$ to be potentially enhanced by factors of three and two, respectively (left panel of Fig.~\ref{fig:Kpinunu}). These enhancements are consistent with the existing measurement of the $K^+\to\pi^+\nu\bar\nu$ decay for most parameter points and allow in particular to reach the central experimental value for the branching ratio of this decay.

Comparing $B$ and $K$ branching ratios such as $Br(B\to\mu^+\mu^-)$ and the short-distance contribution to $Br(K_L\to\mu^+\mu^-)$ as done in the right panel of Fig.~\ref{fig:Kpinunu} one sees that, as outlined above, the effects in the $K$ mode can receive much stronger enhancements than in the $B$ decay mode. For this particular $B$ decay we find the enhancement of the branching ratio to be typically below 15\%.

\begin{figure}
\begin{minipage}{\textwidth}
\begin{center}
\includegraphics[width=.4\textwidth]{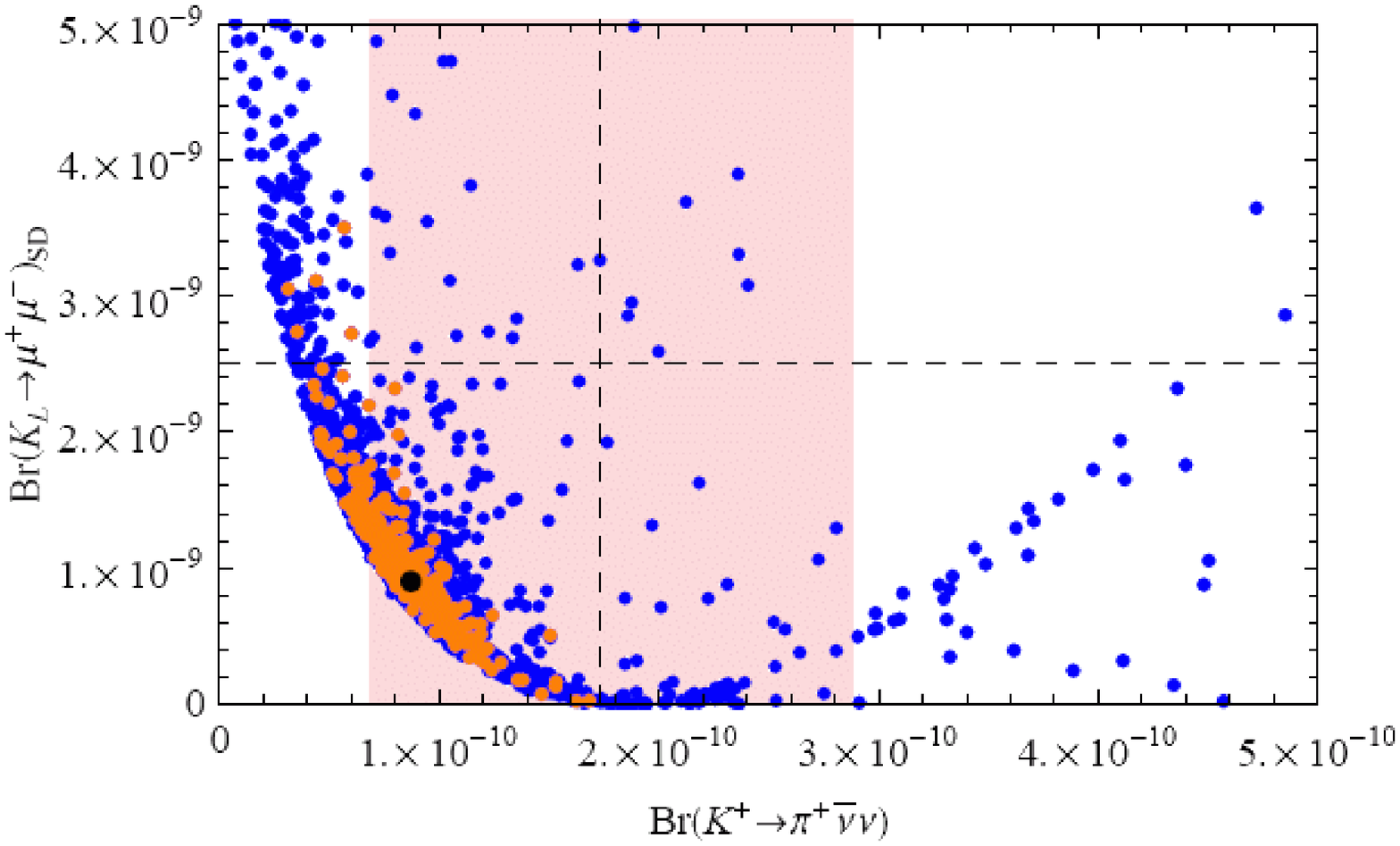}
\hspace{.75cm}
\includegraphics[width=.4\textwidth]{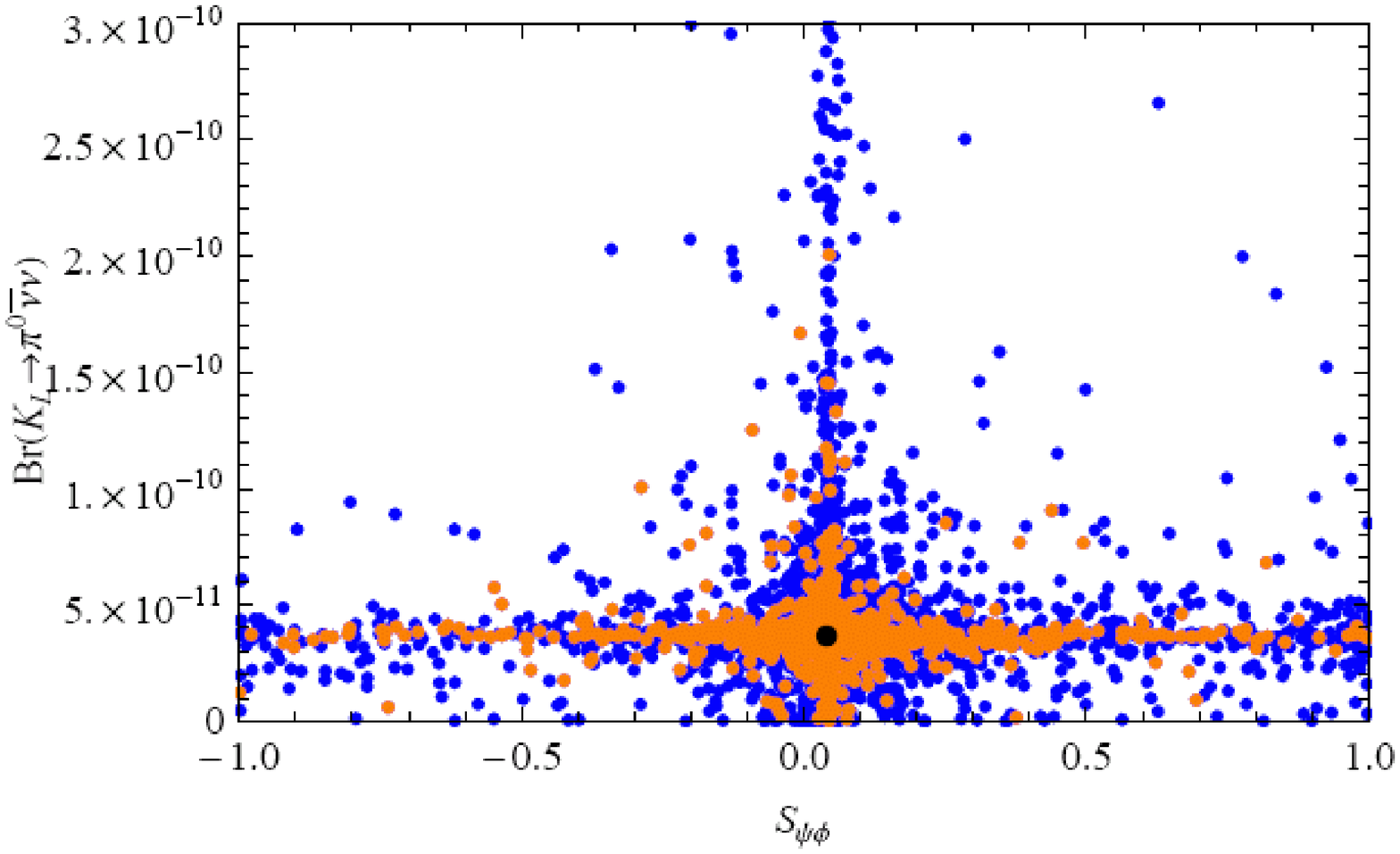}
\end{center}
\end{minipage}
\caption{Left: The short-distance contribution to $Br(K_L\to\mu^+\mu^-)$ plotted against $Br(K^+\to\pi^+\nu\bar\nu)$. Right: $Br(K_L\to\pi^0\nu\bar\nu)$ plotted against $S_{\psi\phi}$.\label{fig:KLmumu}}
\end{figure}

The dominance of the right-handed $Z$ coupling leads to the RS contributions entering $Br(K_L\to\mu^+\mu^-)_\text{SD}$ and $Br(K^+\to\pi^+\nu\bar\nu)$ to enter with opposite sign. This in turn results in an inverse correlation between these two quantities that is shown in the left panel of Fig.~\ref{fig:KLmumu}. This picture is in clear contrast to models where NP enters exclusively or dominantly through left-handed couplings as is the case in the Littlest Higgs model with T-parity (LHT) (for a recent flavour analysis see \cite{Blanke:2009am}). In such models the NP contribution enters both quantities with the same sign and the correlation is found to be roughly linear.

Finally, in the right panel of Fig.~\ref{fig:KLmumu} we contrast the CP violating $\Delta F=2$ and $\Delta F=1$ observables $S_{\psi\phi}$ and $Br(K_L\to\pi^0\nu\bar\nu)$. Both observables can be enhanced significantly beyond their SM values. However it turns out that a simultaneous enhancement of both observables is very unlikely which offers interesting experimental prospects once $S_{\psi\phi}$ has been measured at the LHCb experiment.
\section{Conclusions\label{sec:Conclusions}}
The main results of our analysis of particle-antiparticle mixing \cite{Blanke:2008zb} and rare decays \cite{Blanke:2008yr} can be summarised as follows:
\begin{itemize}
 \item A detailed fine-tuning analysis of $\epsilon_K$ confirms the stringent generic bounds on $M_\text{KK}$ but at the same time shows that the rich flavour structure of the model can accommodate $\epsilon_K$ for mass scales as low as $M_\text{KK}\simeq2.45\,\textrm{TeV}$ with small or moderate fine-tuning
 \item Large effects are possible in $B^0-\bar B^0$ mixing. In particular the CP violating observable $S_{\psi\phi}$ can take values $-1\leq S_{\psi\phi}\leq1$ in the RS model while the SM predicts $S_{\psi\phi}\simeq0.04$ 
 \item For rare decays of $K$ mesons enhancements by factors up to three are possible while the effects in rare $B$ decays are much smaller and typically below 15\%
 \item Simultaneous large enhancements of $S_{\psi\phi}$ and rare $K$ decays are very unlikely. This possibly allows to test the RS model with custodial symmetry experimentally in the foreseeable future.
\end{itemize}

{\bf Acknowledgements}\\
I would like to thank my collaborators Michaela Albrecht, Monika Blanke, Andrzej Buras, Katrin Gemmler, Stefania Gori and Andreas Weiler for the fruitful collaboration that led to the papers this work is based on. This work was partially supported by GRK 1054 of Deutsche Forschungsgemeinschaft.


\begin{thebibliography}{10}

\bibitem{Randall:1999ee}
L. Randall and R. Sundrum,
\newblock Phys. Rev. Lett. 83 (1999) 3370, hep-ph/9905221.

\bibitem{Gherghetta:2000qt}
T. Gherghetta and A. Pomarol,
\newblock Nucl. Phys. B586 (2000) 141, hep-ph/0003129.

\bibitem{Chang:1999nh}
S. Chang et~al.,
\newblock Phys. Rev. D62 (2000) 084025, hep-ph/9912498.

\bibitem{Grossman:1999ra}
Y. Grossman and M. Neubert,
\newblock Phys. Lett. B474 (2000) 361, hep-ph/9912408.

\bibitem{Huber:2003tu}
S.J. Huber,
\newblock Nucl. Phys. B666 (2003) 269, hep-ph/0303183.

\bibitem{Agashe:2004cp}
K. Agashe, G. Perez and A. Soni,
\newblock Phys. Rev. D71 (2005) 016002, hep-ph/0408134.

\bibitem{Agashe:2003zs}
K. Agashe et~al.,
\newblock JHEP 08 (2003) 050, hep-ph/0308036.

\bibitem{Csaki:2003zu}
C. Csaki et~al.,
\newblock Phys. Rev. Lett. 92 (2004) 101802, hep-ph/0308038.

\bibitem{Agashe:2006at}
K. Agashe et~al.,
\newblock Phys. Lett. B641 (2006) 62, hep-ph/0605341.

\bibitem{Cacciapaglia:2006gp}
G. Cacciapaglia et~al.,
\newblock Phys. Rev. D75 (2007) 015003, hep-ph/0607146.

\bibitem{Contino:2006qr}
R. Contino, L. Da~Rold and A. Pomarol,
\newblock Phys. Rev. D75 (2007) 055014, hep-ph/0612048.

\bibitem{Carena:2007ua}
M.S. Carena et~al.,
\newblock Phys. Rev. D76 (2007) 035006, hep-ph/0701055.

\bibitem{Djouadi:2006rk}
A. Djouadi, G. Moreau and F. Richard,
\newblock Nucl. Phys. B773 (2007) 43, hep-ph/0610173.

\bibitem{Bouchart:2008vp}
C. Bouchart and G. Moreau,
\newblock Nucl. Phys. B810 (2009) 66, 0807.4461.

\bibitem{Albrecht:2009xr}
M.E. Albrecht et~al.,
\newblock (2009), 0903.2415.

\bibitem{Blanke:2008zb}
M. Blanke et~al.,
\newblock JHEP 03 (2009) 001, 0809.1073.

\bibitem{Blanke:2008yr}
M. Blanke et~al.,
\newblock JHEP 03 (2009) 108, 0812.3803.

\bibitem{Csaki:2008zd}
C. Csaki, A. Falkowski and A. Weiler,
\newblock JHEP 09 (2008) 008, 0804.1954.

\bibitem{Duling:2009sf}
B. Duling,
\newblock J. Phys. Conf. Ser. 171 (2009) 012061, 0901.4599.

\bibitem{Gori:2009tr}
S. Gori,
\newblock J. Phys. Conf. Ser. 171 (2009) 012062, 0901.4704.

\bibitem{Buras:2009ka}
A.J. Buras, B. Duling and S. Gori,
\newblock (2009), 0905.2318.

\bibitem{delAguila:2000kb}
F. del Aguila and J. Santiago,
\newblock Phys. Lett. B493 (2000) 175, hep-ph/0008143.

\bibitem{Blanke:2009am}
M. Blanke et~al.,
\newblock (2009), 0906.5454.

\end{thebibliography}
\end{document}